\documentclass[aps,floatfix,showpacs,showkeys,preprintnumbers,amsmath]{revtex4}

\usepackage{amssymb}
\usepackage[T1]{fontenc}
\usepackage[latin1]{inputenc}
\usepackage{graphicx}
\usepackage{dcolumn}
\usepackage{bm}

\begin{document}


\title{The effect of the in-medium $\Theta^+$ pentaquark \\ on the kaon optical potential}

\author{Laura Tol\'os}

\affiliation{Institut f\"ur Theoretische Physik $\&$ FIAS. J. W. Goethe-Universit\"at.\\
Max-von-Laue 1. D-60438 Frankfurt am Main, Germany}

\author{Daniel Cabrera}
\affiliation{Departamento de F\'{\i}sica Teórica and
IFIC,
Centro Mixto Universidad de Valencia-CSIC,
Institutos de Investigación de Paterna,
Aptd. 22085,
E-46071 Valencia, Spain}

\author{Angels Ramos, Artur Polls}
\affiliation{Departament d'Estructura i Constituents de la Mat\`eria,
Universitat de Barcelona, \\
Diagonal 647, 08028 Barcelona, Spain
}

\date{\today}

\begin{abstract}

The kaon nuclear optical potential is studied including the effect
of the $\Theta^+$ pentaquark. The one-nucleon contribution is
obtained using an extension of the J\"ulich meson-exchange
potential as bare kaon-nucleon interaction. Significant
differences between a fully self-consistent calculation and the
usually employed low-density $T\rho$ approach are observed. The
influence of the one-nucleon absorption process, $K N \to
\Theta^+$, on the kaon optical potential is negligible due to the
small width of the pentaquark. In contrast, the two-nucleon
mechanism, $K N N \to \Theta^+ N$, estimated from the coupling of
the pentaquark to a two-meson cloud, provides the required amount
of additional kaon absorption to reconcile with data the
systematically low $K^+$-nucleus reaction cross sections found by
the theoretical models.



\end{abstract}



\maketitle

\section{Introduction}
\label{sec:intro}

The study of pentaquarks has become a matter of recent interest
since the discovery by the LEPS collaboration at SPring-8/Osaka
\cite{osaka} of the exotic $\Theta^+$ with strangeness $S=+1$,
which has been confirmed by several other collaborations
\cite{barmin,stepanyan,barth,kubarovsky,airapetian,aleev}. The
possibility of the existence of a narrow baryon resonance of mass
1.53 GeV, width of 15 MeV and quantum numbers $S=+1$, $I=0$ and
$J^P=1/2^+$,  was previously predicted by  the chiral
quark-soliton model of Diakonov et al. \cite{diakonov}. Actually,
the position (1.54 GeV) and width ($<20$ MeV) extracted from the
experiments are compatible with the theoretical prediction.
However, Nussinov \cite{nussinov}, Arndt \cite{arndt} and Gibbs
\cite{gibbs} pointed out that widths larger than a few MeV were
excluded since otherwise the $\Theta^+$ would have been visible in
the $K^+N$ and $K^+d$ data. Similar results were obtained when
comparing the available data on total $KN$ cross sections in the
$I=0$ and $I=1$ channels with the predictions of the J\"ulich
meson-exchange model for the $KN$ interaction
\cite{hoffman,butgen}, extended to incorporate a $\Theta^+$-like
resonance structure \cite{haidenbauer}. In a recent paper, this
extended J\"ulich model has also been used to determine the width
of the $\Theta^+$ resonance through the analysis of the reaction
$K^+ d \rightarrow K^0 pp$ \cite{sibirstev}.

Furthermore, the discovery of the $\Theta^+$ pentaquark with
positive strangeness opens the exciting possibility of producing
exotic $\Theta^+$ hypernuclei. Several authors have proposed that
the $\Theta^+$ would develop in the nuclear medium an attractive
optical potential, the size of which might range, depending of the
mechanism, from a few MeV to a few hundreds of MeV
\cite{miller,cabrera,zhong,shen,navarra}. Therefore, understanding
the in-medium properties of the $\Theta^+$ pentaquark and its
influence on the $KN$ interaction in dense matter and, hence, on
the kaon optical potential is something that deserves being
investigated.

The medium properties of kaons have received a lot of attention
over the last years
\cite{kaplan,waas,li,brown,schaffner,kaiser,oset,nekipelov,korpa}
due to the fact that kaons are not only considered the best probes
to study the dense and hot nuclear matter formed in heavy-ion
collisions, but  also because they probe partial restoration of
the chiral symmetry in dense matter. The $KN$ interaction has been
believed to be smooth since no baryonic resonance with positive
strangeness was allowed to exist and, therefore, the
single-particle potential of kaons has usually been approximated
by the $T \rho$ approximation or low-density theorem with a
repulsion of around 30 MeV for normal nuclear matter density
\cite{kaiser,oset}. Recently, a self-consistent calculation has
been performed \cite{korpa} showing a mass shift of 36 MeV at
normal nuclear matter density. All the theoretical models of the
kaon optical potential based on the $T\rho$ approximation failed
systematically in reproducing $K^+$-nucleus total and reaction
cross sections
\cite{bugg,mardor,krauss,chen,sawafta,weiss,friedman1,friedman2},
underestimating the data by about 10-15\%. Although several
mechanisms where explored, such as swelling of the nucleon, meson
exchange currents, or a smaller mass for the exchanged vector
meson \cite{siegel,brown2,jiang,garcia}, there is at present
no satisfactory solution to this problem. However, a recent work
obtained substantially improved fits to the data by incorporating
the absorption of $K^+$ by nucleon pairs \cite{gal}.

One of the aims of the present work is to revise the validity of
the $T \rho$ approximation to the kaon optical potential, by
performing fully self-consistent calculations using a medium
modified $KN$ effective interaction. In addition, we also
investigate the changes on the kaon optical potential induced by
the presence of the $\Theta^+$ pentaquark in a dense medium.  For
this purpose, we start from an extension of the J\"ulich
meson-exchange model for the $KN$ interaction, which includes the
$\Theta^+$ pentaquark \cite{haidenbauer} as an additional pole
term. From the two models explored in Ref.~\cite{haidenbauer}
giving, respectively, pentaquark widths of 20 MeV and 5 MeV, we
only consider the latter one since it is closer to the upper limit
of various recent quantitative analysis
\cite{arndt,gibbs,sibirstev,polyakov}. We will see that the
mechanism $K N \to \Theta^+$ gives a negligible contribution to
the kaon optical potential due to the small coupling of the
pentaquark to $KN$ states. However, the pentaquark can also
influence the kaon optical potential through the two-body
mechanism $K N N \to \Theta^+ N$, as discussed in Ref. \cite{gal}.
In the present work we perform a microscopic calculation of this
mechanism taking the model of Ref.~\cite{hosaka} for the coupling
of the $\Theta^+$ to a $K \pi$ cloud and allowing the pion to
couple to particle-hole and Delta-hole excitations. We will see
that the effect of this new channel on the kaon optical potential
is appreciable, enhancing the calculated $K^+$ nuclear reaction
cross sections enough to bring them in close agreement with the
experimental data.


\section{The kaon optical potential}
\label{sec:formalism}

In order
to obtain the single-particle potential
 of a $K$ meson embedded in infinite symmetric nuclear matter,
we require the knowledge of the in-medium $KN$
interaction, which will be described by a $G$-matrix.
The medium effects incorporated in this $G$-matrix include
the Pauli blocking on the nucleonic intermediate states as well as
the self-consistent dressing of the $K$ meson and nucleon.

The $KN$ interaction in the nuclear medium is obtained taking,  as
bare interaction $V$, an extension of the meson-exchange J\"ulich
interaction for $KN$ scattering \cite{hoffman,butgen} which
incorporates  the $\Theta^+$ pentaquark as a polar term
\cite{haidenbauer}. The values of the bare pentaquark mass,
$M_{\Theta^+}^0=1545$ MeV, and the bare coupling constant to $KN$
states, $g^0_{KN\Theta^+}/\sqrt{4\pi}= 0.03$, were chosen to
reproduce the observed physical mass and a width of 5 MeV after
solving, in free space, the Lippmann-Schwinger equation which
couples the pentaquark term with the other non-polar terms of the
$KN$ potential.

In a schematic notation, the $G$-matrix reads
\begin{eqnarray}
\langle K N \mid G(\Omega) \mid K N \rangle &&=
\langle K N
\mid V({\sqrt s}) \mid K N \rangle   \nonumber \\
&& \hspace*{-2cm}+ \langle K N \mid V({\sqrt
s }) \mid K N \rangle
\frac {Q_{K N}}{\Omega-E_{K} -E_{N}+i\eta} \langle K N \mid
G(\Omega) \mid K N \rangle \ ,
   \label{eq:gmat1}
\end{eqnarray}
where $\Omega$ is the so-called starting energy, given in the lab
frame, and $\sqrt{s}$ is the invariant center-of-mass energy. In
Eq.~(\ref{eq:gmat1}),  $K$ and $N$  represent, respectively, the
kaon and the nucleon, together with their corresponding quantum
numbers, such as coupled spin and isospin, and linear momentum.
The function $Q_{K N}$ stands for the Pauli operator preventing
scattering to occupied nucleon states  below the Fermi momentum.
Eq.~(\ref{eq:gmat1}) is solved in  a partial wave representation,
including angular momentum channels up to $J=4$. A detailed
description of the method can be found in
Refs.~\cite{Tolos1,Tolos2}.

This $G$-matrix equation has to be considered together with a
prescription for the single-particle energies of kaons and
nucleons in the intermediate states. In the case of kaons, their
single-particle energy is obtained self-consistently from
\begin{equation}
 E_{K}(\vec{q};\rho)=\sqrt{m_{K}^2+\vec{q}\,^2} + {\rm Re}\,U_{K}
(E_{K},\vec{q};\rho) \ , \label{eq:spen}
\end{equation}
where $U_{K}$ is the complex single-particle potential which, in the
Brueckner-Hartree-Fock approach, is given by
\begin{equation}
 U_{K}(E_{K},\vec{q};\rho)= \sum_{N \leq F} \langle K
N \mid
 G_{K N\rightarrow
 K N} (\Omega = E_N+E_{K}) \mid  K N
\rangle \ , \label{eq:self0}
\end{equation}
where the summation over nucleon states extends up to the nucleon
Fermi momentum. The kaon optical potential relates to the kaon self-energy
through $\Pi_K=2 E_K U_K$. The nucleon single-particle energies are taken from
a relativistic $\sigma-\omega$
model with density-dependent scalar and vector coupling
constants \cite{Mach89}. In this model the attraction felt by the zero momentum
nucleons in nuclear matter at saturation density is
of the order of $-$80 MeV.

The kaon optical potential of Eq.~(\ref{eq:self0}) must be
determined self-consistently, since the $KN$ effective interaction
($G$-matrix) depends on the $K$ single-particle energy
Eq.~(\ref{eq:spen}), which in turn depends on the $K$ potential.
We proceed as in Refs.~\cite{Tolos1,Tolos2}, where
self-consistency for the optical potential of the antikaon was
demanded at the quasi-particle energy, a simplification that
proved to be sufficiently good.

\begin{figure}[tb]
\centerline{
     \includegraphics[width=0.5\textwidth]{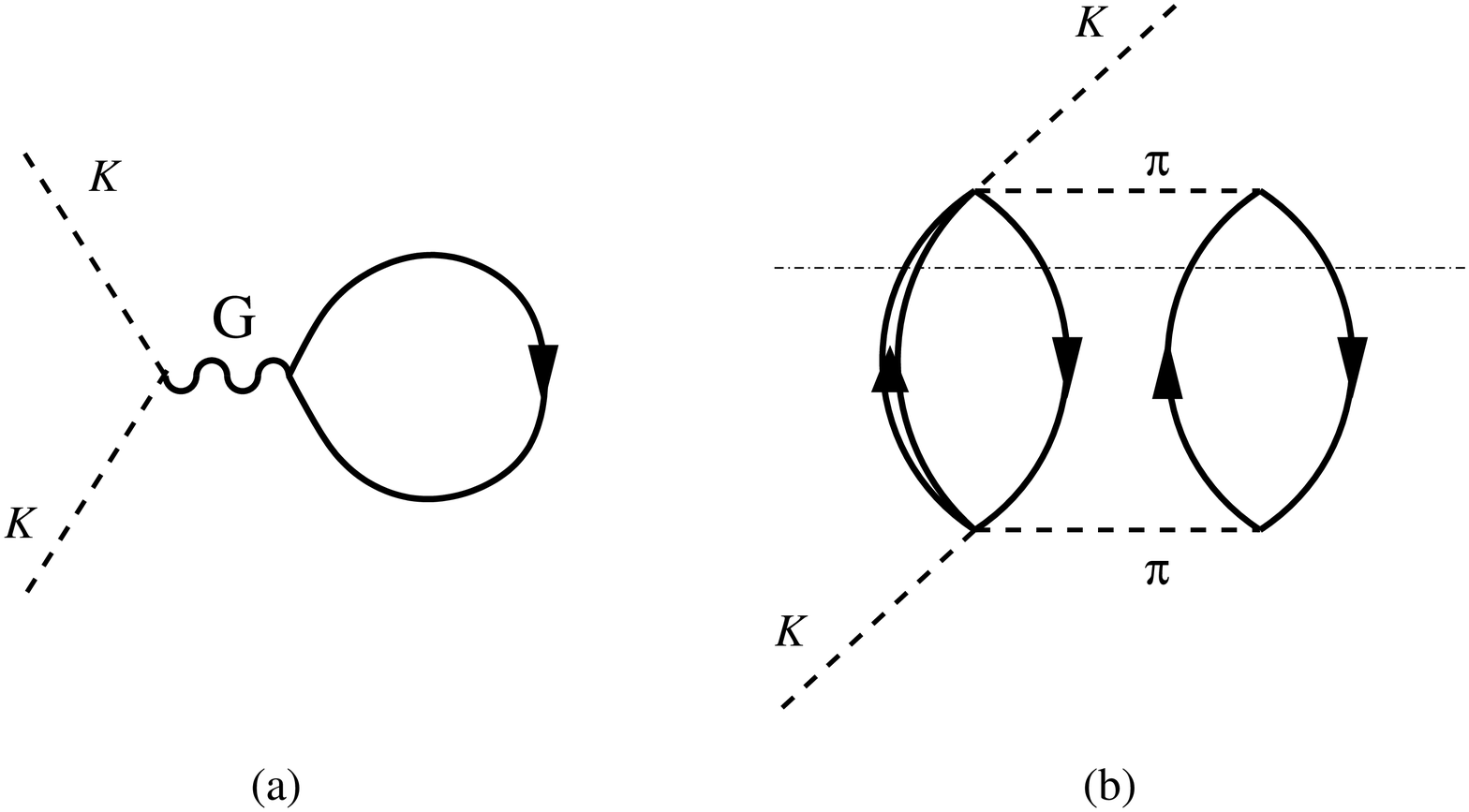}
}
      \caption{One-nucleon (a) and
      two-nucleon (b) contributions to the kaon self-energy.}
        \label{fig:1}
\end{figure}

In a diagrammatic notation, the kaon optical potential obtained
from Eq.~(\ref{eq:self0}) is depicted schematically in
Fig.~\ref{fig:1}(a), with the wiggled line representing the
G-matrix and the solid one a nucleon hole. This diagram implicitly
contains, in the $J=1/2$, $L=1$ $KN$ channel, the effect of the
pentaquark on the kaon optical potential coming from the process
$K N \to \Theta^+$, which is driven by the small value of the
$KN\Theta^+$ coupling constant.

The kaon self-energy may also receive contributions from the
two-nucleon process, $K NN \to \Theta^+ N$, which might be thought
as coming from the mechanism depicted diagrammatically in
Fig.~\ref{fig:1}(b), where a particle-hole (ph) or Delta-hole
($\Delta$h) excitation absorbs the virtual pion emitted at the
$\Theta^+ K\pi N$ vertex. This coupling was evaluated in a study
of the meson cloud effects on the baryon antidecuplet binding in free
space \cite{hosaka} and latter applied in a calculation of the
$\Theta^+$ self-energy in nuclear matter \cite{cabrera}. The
contribution of diagram \ref{fig:1}(b) to the kaon self-energy is
given by
\begin{equation}
\Pi^{2N}_K(q^0,\vec{q};\rho)= i \int\frac{d^4k}{(2\pi)^4}
\left[D_\pi^{(0)}(k)\right]^2
\Pi_\pi(k;\rho) \tilde{U}_\Theta(q,k;\rho) \ ,
\label{eq:selfk2N}
\end{equation}
where $D_\pi^{(0)}(k)$ is the free pion propagator,
$\Pi_\pi(k;\rho)$ stands for the $ph+\Delta h$ contribution to the
pion self-energy and $\tilde{U}_\Theta(q,k;\rho)$ corresponds to
the pentaquark-hole bubble including the $\Theta^+ K\pi N$
vertices, namely
\begin{eqnarray}
\tilde{U}_\Theta(q,k;\rho)=&&  9\, i\sum_{j=S,V}
\int\frac{d^4 p}{(2\pi)^4} \mid
t^{(j)}(k,q,p)
\mid^2 \nonumber \\
 && \left\{\frac{1-n(\vec{p}\,)}{p^0-E_N(\vec{p}\,)+i\varepsilon}
+ \frac{n(\vec{p}\,)}{p^0-E_N(\vec{p}\,)-i\varepsilon} \right\}
\left\{\frac{1}{p^0+q^0-k^0-E_\Theta(\vec{p}+\vec{q}-\vec{k}\,)+i\varepsilon}
\right\} \nonumber \\
\simeq &&- 9 \sum_{j=S,V}  \mid t^{(j)}(k,q) \mid^2
U_\Theta(q-k;\rho) \ , \label{eq:utheta}
\end{eqnarray}
where the isospin factor 9 results from the sum over the processes $K^+ p \to \Theta^+
\pi^+$ and $K^+ n \to \Theta^+ \pi^0$, the quantity $U_\Theta(q-k;\rho)$ stands
for the pentaquark-hole Lindhard function, and the scalar and vector couplings
are given by
\begin{eqnarray}
\mid t^{(S)}(k,q)\mid^2 &=&  -\left(\frac{\tilde{g}}{2f}\right)^2  \left[ 1 +
\frac{M_\Theta}{E_\Theta(\vec{q}-\vec{k}\,)} \right] \nonumber \\
\mid t^{(V)}(k,q)\mid^2 &=&  -\left(\frac{g}{4f^2}\right)^2 \left[
\left( 1 + \frac{M_\Theta}{E_\Theta(\vec{q}-\vec{k}\,)} \right)
(k^0+q^0)^2 + \frac{2}{E_\Theta(\vec{q}-\vec{k}\,)} (\vec{k}^2 -
\vec{q}\,^2) (k^0+q^0) + \right.\nonumber \\
&&\phantom{ -\left(\frac{g}{4f^2}\right)^2 ~}\left. \left( 1 -
\frac{M_\Theta}{E_\Theta(\vec{q}-\vec{k}\,)}\right)(\vec{k} + \vec{q}\,)^2
\right] \left| \frac{m_{K^*}^2}{(q-k)^2-m_{K^*}^2} \right|^2 \ ,
\label{eq:vertices}
\end{eqnarray}
with $f=93$ MeV the pion decay constant, $g= 0.32$ and $\tilde{g}=1.9$ \cite{hosaka}. 
Note that in the last equality of Eq.~(\ref{eq:utheta}) we have
ignored the dependence of the
vertices in the nucleon momentum $\vec{p}$ since it is small in the Fermi sea.

For practical reasons, the diagram of Fig.~\ref{fig:1}(b) is
evaluated replacing, in the integrand of Eq.~(\ref{eq:selfk2N}),
the quantity $\left[D_\pi^{(0)}(k)\right]^2 \Pi_\pi(k;\rho)$ by the full
in-medium pion propagator, $D_\pi(k;\rho)$, which is dressed with a
pion self-energy containing the coupling to particle-hole and
$\Delta$-hole excitations and modified by short range correlations
effects via a Landau-Migdal parameter $g^\prime=0.6$ 
\cite{oset82,carmen88,chiang98}.  Since we are
only interested in the imaginary part of the kaon self-energy and
we keep below the pion production region, we do not need to
subtract the lowest order pion production diagram which is
implicitly implemented by the above mentioned replacement.

\section{Results}
\label{sec:results}

\begin{figure}[tb]
\centerline{
     \includegraphics[width=0.45\textwidth]{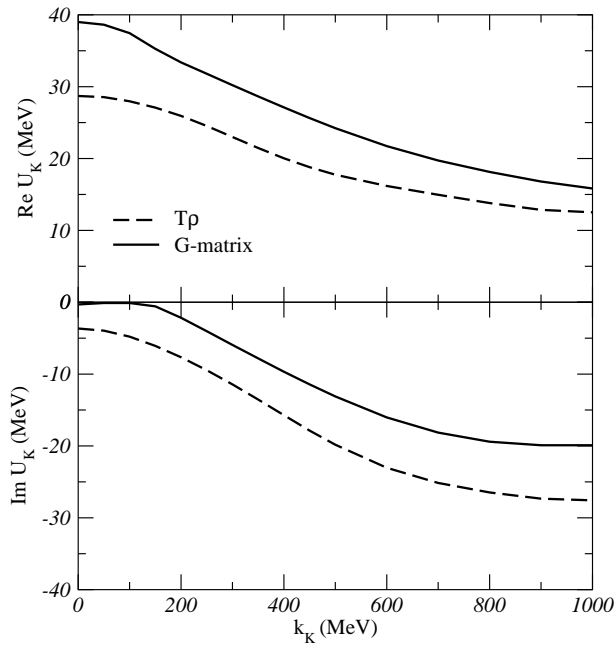}
}
      \caption{Real and imaginary parts of the kaon optical
      potential as functions of the momentum of the kaon at
      normal nuclear matter density for the $T \rho$ approximation (dashed lines)
      and the $G$-matrix calculation (solid lines).}
        \label{fig:graf1}
\end{figure}

We start this section with an analysis of the kaon optical
potential when the $\Theta^+$ pentaquark is not present. In
Fig.~\ref{fig:graf1} we display the real and imaginary parts of
the kaon potential as a function of the  momentum at normal
nuclear matter density, $\rho_0=0.17$ fm$^{-1}$. The solid lines
stand for the self-consistent $G$-matrix calculation and the
dashed lines show the results of the low-density approximation, resulting
from replacing $G$ by the free amplitude $T$ in Eq.~(\ref{eq:self0}).
Our $T\rho$ approach ignores the medium corrections on the $KN$
effective interaction as well as the in-medium single particle potentials
of the $K$ and $N$, but considers the effects of Fermi motion.
This is the reason for having a non vanishing value of
the imaginary part of the $T\rho$ optical potential at zero kaon momentum. For a
momentum of 500 MeV/c, the
size of the imaginary part of the optical potential would be reduced by 5\% if
Fermi motion was disregarded.
We also observe that, at zero momentum, the real
part of the kaon potential in the $T \rho$ approximation is about
10 MeV less repulsive than in the case of the self-consistent
approach which gives an overall repulsive potential of 39 MeV. It
is interesting to note that in Ref.~\cite{korpa}, where
self-consistency was also imposed, a similar repulsion of 10 MeV
with respect to the calculation using the free space amplitudes
was found. Our value of 29 MeV for the kaon optical potential at
zero momentum  in the  $T \rho$ approximation is a few MeV
larger than that in Ref.~\cite{korpa} but very similar to that
obtained with other chiral models \cite{kaiser,oset}.
The imaginary part grows
slowly with increasing momentum and at 400 MeV it has a value of
$-$9 MeV, which would correspond to a width of 18 MeV, slightly
larger again than that found in Ref.~\cite{korpa}.

The results in Fig.~\ref{fig:graf1} demonstrate that medium corrections are
relevant even for the weak, featureless and smoothly energy-dependent $KN$
amplitude. These effects are clearly manifest in the optical potential of the
kaon and, therefore, have an influence on its in-medium properties.

\begin{figure}[thb]
\centerline{
     \includegraphics[width=0.5\textwidth]{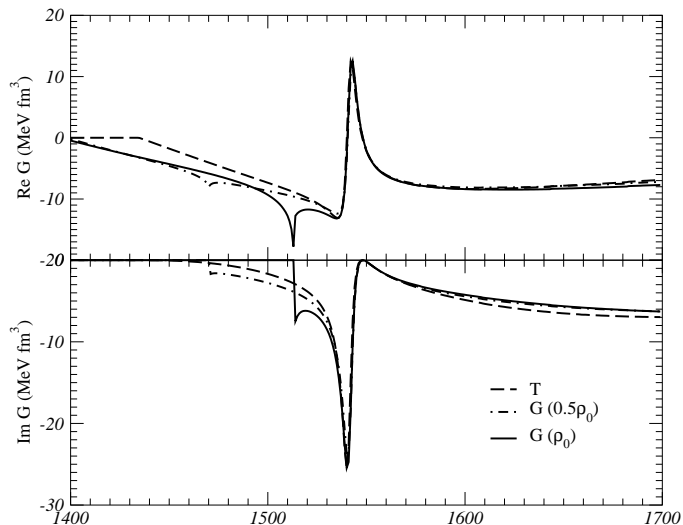}
}
      \caption{Real and imaginary parts of the $G$-matrix for
$L=1$, $J=1/2$ and $I=0$ as functions of the center-of-mass energy
at a total momentum $|\vec{k}_K+\vec{k}_N|=0$ for different
densities considering a $\Theta^+$ pentaquark with a width of 5
MeV. }
        \label{fig:graf2}
\end{figure}

We next study the manifestation of the presence of the $\Theta^+$
pentaquark on the kaon optical potential. For this purpose, we
first show in Fig.~\ref{fig:graf2} the real and imaginary parts of
the $G$-matrix for $L=1$, $J=1/2$ and $I=0$ as a function of the
center-of-mass energy for a total momentum
$|\vec{k}_K+\vec{k}_N|=0$, including a $\Theta^+$ pentaquark with
a width of 5 MeV  in free space. The cases shown correspond to the
free $T$-matrix amplitude (dashed lines), the in-medium $G$-matrix
amplitude for 0.5 $\rho_0$ (dot-dashed lines) and $\rho_0$ (solid
lines). We observe that the structure in the $KN$ amplitude
associated to the presence of the $\Theta^+$ pentaquark barely
changes its position and the width is hardly altered, except for
the fact that it appears distorted by the in-medium modification
of the $KN$ threshold, which at normal nuclear matter density lies
around $1520$ MeV. In spite of the fact that the pentaquark is
included explicitly as an additional polar term in the $KN$
J\"ulich potential, one would not have anticipated this
insensitivity to medium effects. We recall, first, that the
physical position and width of the pentaquark, as seen in the free
space amplitude shown in Fig.~\ref{fig:graf2}, are generated after
the multiple iterations involved in the T-matrix equation.
Secondly, if only the polar pentaquark term of the potential was
iterated, the width would be given from
\begin{equation}
\Gamma = \frac{1}{\pi}\left( \frac{g^0_{KN\Theta^+}}{2m_N}\right)^2
\frac{m_N}{M_{\Theta^+}}q_{\rm on}^3 \ ,
\end{equation}
and, taking the bare value for the coupling constant quoted above,
one would obtain a value around 1.2 MeV, four times narrower than
it ends up being when the complete potential is iterated in the
Lippmann-Schwinger equation. In other words, in the extended
J\"ulich model the width of the pentaquark is largely generated
from the interferences between the polar and non-polar terms of
the interaction. Since medium effects affect the size of these
interferences (by modifying the phase space of intermediate $KN$
states) one in principle would have expected larger in-medium
corrections on the width of the pentaquark. Changes are drastic only for higher
densities, where the in-medium $KN$ threshold moves to energies
higher than the position of the pentaquark, and the width obviously
drops to zero.

\begin{figure}[tb]
\vspace{1cm}
\centerline{
     \includegraphics[width=0.6\textwidth]{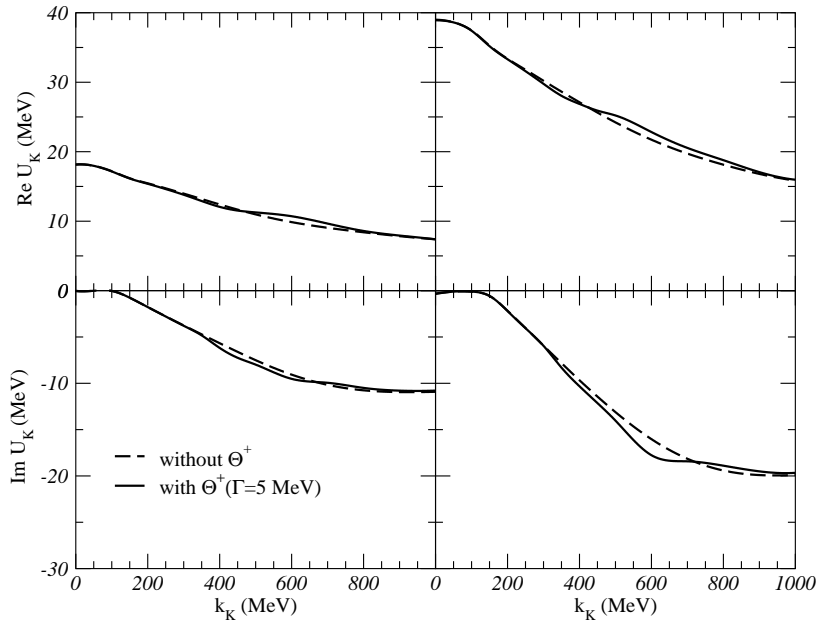}
}
      \caption{\small Real and imaginary parts of the kaon optical
       potential as functions of the momentum of the kaon for
       $0.5$ $\rho_0$ (left panels) and $\rho_0$ (right panels), without
       $\Theta^+$ (dashed lines) and
       including a $\Theta^+$ resonance with a width of 5
       MeV (solid lines). }

        \label{fig:graf3}
\end{figure}

Once we have obtained the in-medium $KN$ amplitude including the
effect of the $\Theta^+$ pentaquark, we can proceed to examine the
consequences of the existence of this pentaquark on the optical
potential of kaons. In Fig.~\ref{fig:graf3} the real and imaginary
parts of the kaon optical potential are displayed as functions of
the momentum for $0.5$ $\rho_0$ (left panels) and $\rho_0$ (right
panels). The self-consistent kaon potential without $\Theta^+$ is
shown by the dashed lines. The solid line represents the optical
potential when the $K N \to \Theta^+$ mechanism is included.
Although some changes are seen for momentum values larger than 300
MeV/c, the effect is practically negligible and this is tied to
the small value of the $KN\Theta^+$ coupling constant.

\begin{figure}[tb]
\centerline{
     \includegraphics[width=0.45\textwidth]{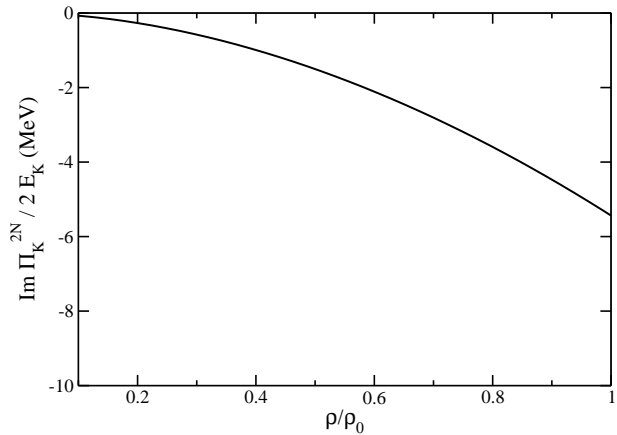}
}
      \caption{\small Imaginary part of the $2N$ contribution to
      the kaon optical
       potential as function of the density for a kaon momentum of
       500 MeV/c.}

        \label{fig:selfk2N}
\end{figure}

We next present in Fig.~\ref{fig:selfk2N} the contribution to the
 imaginary part of the
the kaon optical potential coming from the $2N$ mechanism as a
function of the nuclear density for a kaon momentum of 500 MeV/c.
These results exhibit the expected $\rho^2$ dependence and have a
significant size. At normal nuclear matter density, the
contribution of the two-nucleon mechanism to the imaginary part of
the kaon optical potential for a momentum of 500 MeV/c is almost 
half of that corresponding to one-nucleon reactions.

Following Ref.~\cite{oset85}, we calculate absorption and reaction cross
sections, $\sigma_{\rm abs}$ and $\sigma_{R}$, from the expression:
\begin{equation} 
\sigma= \int d^2b \left[ 1 - {\rm exp}\left( -
\int_{-\infty}^\infty -\frac{1}{q} {\rm Im}\, \Pi(q;\rho(\vec{b},z)) dz \right)
\right] \ , 
\label{eq:cross1}
\end{equation} 
using, respectively, the two-nucleon component of
the kaon self-energy, $\Pi^{2N}_K$, or the total self-energy containing, in
addition, the contribution of the one-nucleon processes. 
The cross section is thus obtained as the integral over the impact parameter of
one minus the probability that the kaon crosses the nucleus without reacting (in
case of $\sigma_{R}$) or without being absorbed (in case of $\sigma_{\rm abs}$).
We note that, in the latter case, we do not remove from the flux the kaons that
undergo quasielastic collisions since they can still be absorbed. Although the
eikonal formalism assumes the quasielastic scattered kaons to keep moving in
the forward direction, it is still a reasonably good approximation for
inclusive observables as is the case of the cross sections calculated here. An upper
bound of the effect of distortions on the absorption cross
sections would be obtained from the expression:
\begin{equation} 
\sigma_{\rm abs}= \int d^2b\, dz\, {\rm exp}\left[ -
\int_{-\infty}^z -\frac{1}{q} {\rm Im}\, \Pi(q;\rho(\vec{b},z^\prime)) dz^\prime
\right] (-) \frac{1}{q} {\rm Im}\, \Pi_{\rm abs}(q;\rho(\vec{b},z)) \ , 
\label{eq:cross2}
\end{equation} 
with $\Pi$ being the total self-energy,
which removes from the flux the kaons that have
suffered any type of reaction before reaching the absorption point $z$. 
This would be more
in line with the Distorted Wave Impulse Approximation-like expression used in
Ref.~\cite{gal}.

\begin{figure}[tb]
\centerline{
     \includegraphics[width=0.45\textwidth]{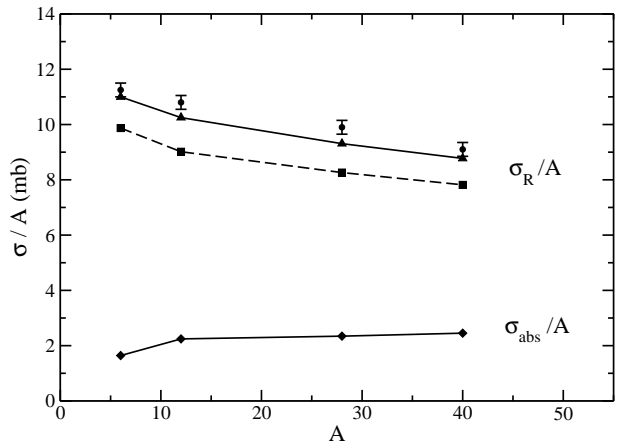}
}
      \caption{\small Absorption and reaction cross sections per nucleon for a kaon
      laboratory
      momentum of 488 MeV/c, from a $G\rho$ kaon
      optical potential (dashed line) and including, in addition,
      the 2N-absorption mechanism (solid line).
      Data for the reaction cross sections are taken from the
      analysis of \protect\cite{friedman2}.}

        \label{fig:sigma}
\end{figure}

In Fig.~\ref{fig:sigma}  the calculated cross sections per nucleon in $^6$Li,
$^{12}$C, $^{28}$Si and $^{40}$Ca using Eq.~(\ref{eq:cross1})
are compared to experimental data. The dashed
line is obtained with the $G\rho$ kaon optical potential of
Eq.~(\ref{eq:self0}) and underestimates the data by about 15\%. By inspecting
Fig.~\ref{fig:graf1} one would then expect the $T\rho$ model 
to properly describe or even overestimate the data, in apparent 
contradiction to the previous theoretical
calculations which underestimate data by 10-15\%. We note, however, that 
we are not using here the experimental amplitudes but those obtained with the J\"ulich
model I of Ref.~\cite{hoffman}. This model reproduces the threshold scattering
observables, but the $I=0$ and $I=1$ cross sections at 500 MeV/c and higher are slightly 
larger than the experimental ones. 
The absorption cross sections per nucleon obtained from the
$2N$ mechanism are about 2-3 mb, right below the upper bound of 3.5 mb
established in Ref.~\cite{gal} from a phenomenological analysis including $2N$
absorptive effects. The use of the alternative Eq.~(\ref{eq:cross2}) produces
lower values of the absorption cross sections, slightly below 2 mb. 
The reaction cross sections per nucleon obtained with the
complete imaginary part of the kaon self-energy, including both the one- and
two-nucleon processes, lie very close to the experimental data. Clearly, our
model for the two-nucleon kaon absorption mechanism provides the required
strength to bring the reaction cross sections in agreement with experiment,
thereby giving a  possible answer to a  long-standing anomaly in the physics of
kaons in nuclei.

We would also like to comment on the possibility that the
in-medium pentaquark develops, through its interaction with
nucleons, an attractive potential, which might range from a few
MeV to an astonishing value of a few hundreds of MeV
\cite{miller,cabrera,zhong,shen,navarra}. This attraction can be
thought as coming from a strong coupling of the pentaquark to $K^*
N$ states. The J\"ulich model does not incorporate such a coupling
and, as a consequence, it is not included in the corresponding
one-nucleon contribution to the kaon optical potential displayed in
Fig.~\ref{fig:graf3}. Actually, the coupling to $K^*N$ states
would have moved to smaller momentum values the range of influence of
the pentaquark on the kaon self-energy, but its effect would remain 
negligible since it is driven by
the small value of the $\Theta^+ KN$ coupling constant. The
two-nucleon mechanism partly includes the $\Theta^+ K^* N$
coupling since the two-meson cloud in the $\Theta^+ K \pi N$
vertex reconstructs a $K^*$ meson, as discussed in
Ref.~\cite{hosaka} and illustrated by the presence of the
vector-meson form factor in Eq.~(\ref{eq:vertices}). In addition,
and for consistency, we should dress the pentaquark in diagram 1(b) by adding an
attractive potential to its energy. However, a similar binding
should be considered for the nucleon and both effects largely
compensate each other.  In any case, the possible induced error
certainly lies within the uncertainties  of our model for the
two-nucleon absorption mechanism, which we estimate to be within
30\%.

\section{Conclusions}
\label{sec:conclusions}


We have performed a microscopic self-consistent calculation of the
single-particle potential of a $K$ meson embedded in symmetric
nuclear matter, using the meson-exchange J\"ulich $K N$
interaction. We have investigated the differences between the full
self-consistent calculation of the $K$ optical potential and the
$T \rho$ approximation. The medium modifications on the
$KN$ amplitude affect the value of the $K$ optical
potential.
While the $T \rho$ approach gives a repulsive potential of 29 MeV
at zero momentum, the full self-consistent calculation increases
the repulsion up to 39 MeV, in line with what is observed in
the work of Ref.~\cite{korpa}, where self-consistency was also implemented.

We have also studied the effect on the kaon optical potential of
the existence of the $\Theta^+$ pentaquark in a dense medium. We
first obtain an in-medium $KN$ effective interaction starting from
an extension of the meson-exchange potential of the J\"ulich
group, which includes the $\Theta^+$ resonance. Neither the
position nor the width of the pentaquark change appreciably in the
medium, as long as there is available $KN$ phase space to decay
into. We note, however, that the pentaquark would have developed
a strong binding in the medium if the coupling to $K^* N$ states
had been explicitly included in the J\"ulich model.

The effect of the $\Theta^+$ on the kaon optical potential coming
from the one-nucleon process $K \to \Theta^+ N$ is negligible due
to a very small value of the $\Theta^+ K N$ coupling constant. In
contrast, the two-nucleon absorption mechanism, $K NN \to \Theta^+
N$, contributes significantly to the imaginary part of the kaon
optical potential, being almost half of the $G\rho$ value at normal
nuclear matter density.

The new two-nucleon mechanism calculated in this work produces
$K^+$ nuclear absorption cross sections per nucleon of about 2-3
mb. In addition, the reaction cross sections per nucleon are
increased by 10-15\%  with respect to the $G\rho$ values, and turn
out to be practically in agreement with the experimental data,
thereby confirming the expectations of a recent phenomenological
analysis \cite{gal}.

\section*{Acknowledgments}


We warmly thank
J. Haidenbauer for kindly providing us with the extended J\"ulich
code and for helping us in making it work properly. 
We also thank L. Roca and M.J. Vicente-Vacas for fruitful comments and discussions. 
L.T. wishes to acknowledge support from the
Alexander von Humboldt Foundation. D.C. acknowledges support from the
Ministerio de Educaci\'on y Ciencia. This work is partly supported
by DGICYT contract BFM2002-01868, the Generalitat de Catalunya
contract SGR2001-64, and the E.U. EURIDICE network contract
HPRN-CT-2002-00311. This research is part of the EU Integrated
Infrastructure Initiative Hadron Physics Project under contract
number RII3-CT-2004-506078.


\end{document}